\PassOptionsToPackage{table}{xcolor}

\documentclass[acmlarge]{acmart}

\usepackage{booktabs}
\usepackage{multirow}

\newcolumntype{C}[1]{>{\centering\arraybackslash}p{#1}}
\newcolumntype{L}[1]{>{\raggedright\arraybackslash}p{#1}}
\definecolor{lightgray}{gray}{0.9} 

\AtBeginDocument{%
  }

\begin{document}

\title{Less is More: Simplifying Network Traffic Classification Leveraging RFCs}

\author{Nimesha Wickramasinghe}
\email{n.wickramasinghe@unsw.edu.au}
\affiliation{%
\department{School of Computer Science and Engineering}
  \institution{University of New South Wales}
  \city{Sydney}
  \state{NSW}
  \country{Australia}
}

\author{Arash Shaghaghi}
\email{a.shaghaghi@unsw.edu.au}
\affiliation{%
\department{School of Computer Science and Engineering}
  \institution{University of New South Wales}
  \city{Sydney}
  \state{NSW}
  \country{Australia}
}

\author{Elena Ferrari}
\email{elena.ferrari@uninsubria.it}
\affiliation{%
\department{Department of Theoretical and Applied Science}
  \institution{University of Insubria}
  \city{}
  \state{Varese}
  \country{Italy}
}

\author{Sanjay Jha}
\email{sanjay.jha@unsw.edu.au}
\affiliation{%
\department{School of Computer Science and Engineering}
  \institution{University of New South Wales}
  \city{Sydney}
  \state{NSW}
  \country{Australia}
}

\begin{abstract}
The rapid growth of encryption has significantly enhanced privacy and security while posing challenges for network traffic classification. Recent approaches address these challenges by transforming network traffic into text or image formats to leverage deep-learning models originally designed for natural language processing, and computer vision. However, these transformations often contradict network protocol specifications, introduce noisy features, and result in resource-intensive processes. To overcome these limitations, we propose \textit{NetMatrix}, a minimalistic tabular representation of network traffic that eliminates noisy attributes and focuses on meaningful features leveraging RFCs (Request for Comments) definitions. By combining NetMatrix with a vanilla XGBoost classifier, we implement a lightweight approach, \textit{LiM ("Less is More")} that achieves classification performance on par with state-of-the-art methods such as ET-BERT and YaTC. Compared to selected baselines, experimental evaluations demonstrate that LiM improves resource consumption by orders of magnitude. Overall, this study underscores the effectiveness of simplicity in traffic representation and machine learning model selection, paving the way towards resource-efficient network traffic classification.
\end{abstract}

\maketitle

\section{Introduction}
The rapid growth of encrypted network traffic has significantly enhanced user privacy and security. However, it poses challenges for network traffic classification (NTC), a critical task for network management, security monitoring, and quality of service provisioning. Traditional classification methods that rely on inspecting packet payloads are rendered ineffective by encryption protocols like TLS 1.3 [RFC-8446], which are designed to prevent the extraction of meaningful patterns from encrypted payloads \cite{foundation_et_bert}.

Although many novel approaches are proposed for NTC, they often make several assumptions and employ methodologies that do not align with modern encryption specifications. Below, we present the resultant challenges (C) in NTC by comparing widespread assumptions with modern-day RFCs.

\textbf{\textit{C1.}Assumption of implicit patterns in encrypted payloads}: Many models operate under the premise that encrypted payloads contain exploitable patterns for classification \cite{foundation_et_bert, foundation_yatc_full}. This contradicts the TLS 1.3 specification [RFC-8446], which ensures that the same plaintext will always produce different ciphertexts due to the use of Authenticated Encryption with Associated Data (AEAD) ciphers and unique initialization vectors (IVs).

\textbf{\textit{C2.}Fixed-Size homogeneous representation of payloads}: Studies often truncate or pad payloads to create fixed-size inputs. This approach ignores TLS 1.3's acknowledgment that encrypted packet lengths and timing are susceptible to traffic analysis attacks [RFC-8446]. Padding or truncation does not effectively exploit this vulnerability and may obscure valuable size-related artifacts essential for classification.

\textbf{\textit{C3.}Inclusion of header without justification}: Including header fields such as IP Identification (IP-ID) [RFC-6274], IP header checksum [RFC-791], sequence and acknowledgment numbers [RFC-9293], and TCP options timestamps [RFC-7323] introduces noise. These fields often contain pseudo-random values initialized per session, adding unnecessary variability to the data.

\textbf{\textit{C4.}Representation of network traffic as text or images}: Transforming network traffic into textual or visual formats to leverage semantic, syntactic, or spatial relationships using  foundation models (e.g., BERT, vision transformers) can hinder the inherent structure of network protocols. Unlike natural language tokens or image pixels, network header attributes typically lack interdependencies, making such representations less effective.

\textbf{\textit{C5.}Reliance on TCP/TLS handshake packets}: Some classifiers \cite{foundation_yatc_full} depend on initial handshake packets for classification. However, mechanisms like TCP connection reuse [RFC-9293], TLS 1.3's 0-RTT data transmission, and session resumption via Pre-Shared Keys (PSKs) [RFC-8446] allow communication without frequent handshakes. Consequently, classifiers that rely on handshake packets may fail in these common scenarios. 

These challenges underscore the need for a practical and efficient approach to NTC. To bridge this gap, we propose \textit{NetMatrix}, a novel tabular representation that focuses on capturing meaningful features of network traffic leveraging RFC specifications (\S \ref{subsec:netmatric}). NetMatrix, paired with a vanilla XGBoost classifier, referred to as \textit{LiM} \footnote{Source code: https://github.com/nime-sha256/LiM} (Less is More) (\S \ref{subsec:lim}), achieves performance comparable to resource-intensive deep-learning models while maintaining significantly lower computational overhead (\S \ref{subsubsec:classification}). Moreover, our approach demonstrates the ability to handle high volumes of requests, adapt to new traffic classes within seconds, and operate with enhanced memory and energy efficiency (\S \ref{subsubsec:scalability}).







\section{Literature Review}
Encrypted NTC has been a well-researched topic, with numerous models proposed to address the complexities of encryption protocols like TLS 1.3. Notably, methods such as ET-BERT \cite{foundation_et_bert} and YaTC \cite{foundation_yatc} have leveraged sophisticated deep learning techniques to classify encrypted traffic effectively.

ET-BERT \cite{foundation_et_bert} applies Bidirectional Encoder Representations from Transformers (BERT), originally developed for natural language processing (NLP). The method assumes that encrypted payloads contain implicit patterns that can be exploited for classification ($C1$). The best performing version of ET-BERT (ET-BERT Packet-wise classifier) truncates payloads and represent network traffic using 128 bytes per packet ($C2$). This method involves converting binary raw data (i.e. header and encrypted payload) into sequences of tokens. The consequent transformation aims to capture semantic and syntactic relationships within transformed network traffic, analogous to words and sentences in natural language ($C4$).

Similarly, YaTC \cite{foundation_yatc} proposes representing network traffic as images and employs Masked Autoencoders (MAEs) for classification. Each network session is represented by 1,600 bytes of raw data, arranged into two-dimensional matrices to form images. This approach attempts to exploit spatial relationships within the transformed network traffic ($C4$). For traffic representation, YaTC relies on the first 5 packets of a network traffic session ($C5$).

While ET-BERT and YaTC are known for their great performance \cite{netbench, yatc_etbert_best_1}, other studies have also explored various aspects of encrypted NTC using raw packet data without justifications \cite{tfe-gnn, foundation_flow_mae} ($C3$). However, most existing models for encrypted NTC often face challenges due to their: (1) reliance on encrypted payload content, (2) improper handling of data representations, (3) inclusion of noisy header information, (4) misalignment with protocol specifications, and (5) dependency on handshake packets. 

Our work distinguishes itself by fundamentally rethinking the assumptions in the context of modern encryption standards. By aligning our methodology with RFCs and focusing on meaningful, structured data representations, we aim to provide a more effective and lightweight solution to encrypted network traffic classification.

\section{Methodology}
In this section, we present our proposed methodology for efficient encrypted network traffic classification. In \S \ref{subsec:netmatric}, we introduce \textit{NetMatrix}, the representation that captures the essential features of network traffic according to RFCs. Building upon this representation, in \S \ref{subsec:lim}, we explain the machine learning model selection for experimental evaluation. 


\subsection{NetMatrix}
\label{subsec:netmatric}
NetMatrix is a concise tabular representation of network traffic designed to overcome the limitations of existing NTC methods. The proposed representation is built upon the following principles aligning to modern-day RFCs such as, RFC-8446, RFC-2328, RFC-4271, RFC-6247, RFC-791, RFC-9293, RFC-7323:

\textbf{Exclusion of Encrypted Payloads}: Recognizing that encrypted payload content lacks exploitable patterns as per TLS 1.3 specifications [RFC-8446], we entirely exclude it from NetMatrix to address ($C1$).

\textbf{Utilization of Encrypted Payload Length and Timing}: Acknowledging that TLS 1.3 highlights the susceptibility of encrypted packet lengths and timings to traffic analysis attacks [RFC-8446], we incorporate these attributes and solve ($C2$). Specifically, we use the total length field from the IP header and calculate the inter-arrival times between packets.

\textbf{Reduction of Representation Size}: By excluding noisy, session-specific header fields—such as IP Identification (IP-ID), checksums, sequence numbers, and TCP timestamps, we focus on stable, meaningful header attributes aligning to RFCs. This reduction minimizes noise and enhances classification performance by concentrating on relevant data, resolving ($C3$).

\textbf{Preservation of Inherent Network Traffic Schema}: Instead of transforming network traffic into text or images, which can obscure protocol structures, NetMatrix maintains the inherent schema by using a tabular format as a remedy for ($C4$). This representation preserves the relationships defined by network protocols.

\textbf{Reliance on Packets Containing Encrypted Data}: To ensure practicality, we base our classification on packets that contain encrypted payloads. This approach remains effective even when handshake packets are unavailable as a solution to ($C5$).

NetMatrix utilizes only three key attributes extracted from network packets to represent a session:

\textbf{Total Length (IP Header)}: The total length field in the IP header indicates the size of the entire IP packet, including both the header and the payload.

\textbf{Time-to-Live (TTL) (IP Header)}: The TTL field represents the maximum number of hops a packet can traverse before being discarded. While TTL values can vary due to routing changes, they are generally consistent for packets taking the same route between two endpoints. Routing protocols aim for minimal overhead, so minor route fluctuations have limited impact on the TTL upon arrival [RFC-2328, RFC-4271]. Therefore, TTL can provide useful information about network path characteristics.

\textbf{Inter-Arrival Time}: This attribute measures the time difference between two consecutive packets.

NetMatrix represents each network traffic session (i.e., a bidirectional flow between two endpoints) by extracting the aforementioned attributes from five consecutive packets containing encrypted payloads. For each packet, we collect: Total Length (2 bytes), TTL (1 byte), Inter-Arrival Time (calculated, 3 bytes). This results in a compact representation of 30 bytes per session, significantly reducing the data volume compared to existing methods. For instance, ET-BERT uses 620 bytes, and YaTC requires 1,600 bytes to represent a session. 


\subsection{LiM}
\label{subsec:lim}

To complement the efficient representation provided by NetMatrix, we employ a vanilla XGBoost classifier \cite{xgboost}. As highlighted by \cite{xgboost_tabular_data}, tree-based models, such as XGBoost, are particularly adept at learning irregular patterns in the target function, and demonstrate superiority on tabular data. This insight justifies our selection of the XGBoost classifier in this study. Its inductive biases make it inherently well-suited for the structured nature of NetMatrix, enabling efficient and accurate classification without requiring the computational resources demanded by deep-learning models.

We refer to the integration of NetMatrix with the XGBoost classifier as \textit{LiM (Less is More)}. By utilizing LiM, we demonstrate that a simple machine learning model can achieve performance comparable to state-of-the-art deep learning approaches, such as ET-BERT and YaTC, without the need for complex architectures or extensive computational resources.

\section{Experimental Evaluation}
To validate the effectiveness and efficiency of our proposed approach, we conduct an experimental evaluation focusing on encrypted traffic adhering to TLS 1.3. Our evaluation consists of two main components: (1) assessing classification performance, and (2) analyzing resource consumption.

We utilize the CSTNET-TLS1.3 dataset \cite{foundation_et_bert}, introduced in the ET-BERT study, which comprises network traffic encrypted with TLS 1.3. This dataset is well-suited for our evaluation as it reflects modern encryption protocols and the associated challenges in NTC. To address the class imbalance issue, we randomly select 10 classes from the dataset, each containing more than 400 samples. For an unbiased evaluation, we provide 400 samples per class to each classifier. This equal provision ensures that no class dominates the training process, and facilitates a fair assessment of resource consumption.

\subsection{Experimental Setup}
\subsubsection{Hardware Configuration}: The experiments were conducted on a machine equipped with dual Intel Xeon Silver 4208 CPUs operating at a base clock of 2.10 GHz, with a maximum clock speed of 3.20 GHz. Each processor features 8 cores per socket, for a total of 16 physical cores and 32 threads. The system is equipped with 125 GiB of RAM. For GPU acceleration, the machine houses four NVIDIA Tesla T4 GPUs, each with 16 GiB of dedicated memory, summing up to a total GPU memory capacity of 64 GiB. The GPUs have a maximum power draw capacity of 70W each.

\subsubsection{Software Configuration}: The system runs on Ubuntu 24.04.1 LTS (Noble) as the operating system. The software environment includes Python 3.13.0 as the programming language and the XGBoost 2.1.2 library for implementing and training LiM. \\

For a comparative analysis, we also evaluated state-of-the-art models ET-BERT and YaTC using their recommended configurations. These models were executed on the same experimental setup to ensure a fair and consistent comparison across different classifiers

\subsection{Evaluation Metrics}
The evaluation encompassed both classification performance and resource consumption to provide a comprehensive assessment of each selected model's effectiveness. For classification performance, we employed standard metrics including \textit{accuracy}, \textit{precision}, \textit{recall}, and \textit{F1 score}.

Regarding resource consumption, we evaluated \textit{latency}, \textit{memory usage}, \textit{energy consumption}, and \textit{throughput}. \textit{Latency} was measured as the time taken by the classifier to process a single input during training, calculated by dividing the total time by the number of training samples. The rest of the matrices were measured during the inference phase. \textit{Memory usage} referred to the amount of RAM+GPU consumed, an important factor for deployment on devices with limited resources. \textit{Energy consumption} represented the amount of energy used by the GPU, reflecting the operational cost and sustainability of the model. \textit{Throughput} indicated the number of inferences the classifier could make per second, providing insight into its ability to handle high-volume traffic, which is essential for network environments with heavy loads. 

\subsection{Results and Analysis}

\subsubsection{Classification Performance} \label{subsubsec:classification} \hfill\\
The results, as presented in Table \ref{tab:1_classification_performance}, demonstrate the performance of ET-BERT, YaTC, and the proposed LiM classifier.

\begin{table}[tb]
\renewcommand{\arraystretch}{1.1}

    \caption{Classification Performance Comparison}
    \label{tab:1_classification_performance}
    
    \centering

    \resizebox{0.6\columnwidth}{!}{
        
        \begin{tabular}{l*{4}{C{1.9cm}}} 
            
            \toprule
            
                \multicolumn{1}{c}{\textbf{Model}} & 
                \multicolumn{1}{c}{\textbf{Accuracy}} & 
                \multicolumn{1}{c}{\textbf{Recall}} & 
                \multicolumn{1}{c}{\textbf{Precision}} & 
                \multicolumn{1}{c}{\textbf{F1 Score}} \\ 
            
            \midrule
            
                ET-BERT \cite{foundation_et_bert}  & 
                 0.568 & 
                 0.568 & 
                 0.581 &
                 0.569 \\

                 \cmidrule(lr){1-5}

                YaTC \cite{foundation_yatc}  & 
                 0.963 & 
                 0.963 & 
                 0.964 &
                 0.963 \\

                 \cmidrule(lr){1-5}

                \rowcolor{lightgray}        
                LiM  & 
                 0.942 & 
                 0.942 & 
                 0.943 &
                 0.942 \\
                
            \bottomrule



                    
                
        
        \end{tabular}
    }
\end{table}

ET-BERT achieves a classification accuracy of 0.568, with recall and F1 score also at 0.568 and 0.569, respectively, while precision is marginally higher at 0.581. The underperformance is likely due to the constrained training setup (400 samples per class) provided for a fair evaluation across all models.

In contrast, the YaTC model performs significantly better, achieving an accuracy and F1 score of 0.963.

The proposed LiM classifier, when paired with the NetMatrix representation, achieves competitive results with an accuracy of 0.942 and an F1 score of 0.942. Precision and recall are also closely matched at 0.943 and 0.942, respectively, underscoring the model's robustness across all evaluation metrics.

While LiM slightly trails YaTC in overall performance, its results highlight the strength of a simple yet effective approach. The combination of the NetMatrix representation and the XGBoost classifier demonstrates that high classification performance can be achieved without reliance on complex architectures, offering a compelling balance between accuracy and simplicity.

\subsubsection{Resource Consumption} \label{subsubsec:scalability} \hfill\\
The lightweight XGBoost classifier, when paired with the NetMatrix representation (LiM), demonstrates less resource consumption across all evaluated metrics, as shown in Table \ref{tab:2_resource_consumption}

\begin{table}[tb]
\renewcommand{\arraystretch}{1.1}

    \caption{Scalability Comparison}
    \label{tab:2_resource_consumption}
    
    \centering

    \resizebox{0.6\columnwidth}{!}{
        
        \begin{tabular}{l*{4}{C{1.9cm}}} 
            
            \toprule
            
                \multicolumn{1}{c}{\centering\textbf{Model}} & 
                \textbf{Latency} \small{(Sec/Sample)} & 
                \textbf{Memory} \small{(MiB)} & 
                \textbf{Energy} \small{(Watt)} & 
                \textbf{Throughput} \small{(Sample/Sec)} \\ 
            
            \midrule
            
                ET-BERT \cite{foundation_et_bert}  & 
                 0.4328 & 
                 15,189.48 & 
                 109 &
                 164.91 \\

                 \cmidrule(lr){1-5}

                YaTC \cite{foundation_yatc}  & 
                 0.1342 & 
                 5,352.25 & 
                 26 &
                 636.62 \\

                 \cmidrule(lr){1-5}

                \rowcolor{lightgray}        
                LiM  & 
                 0.0005 & 
                 196.38 & 
                 0 &
                 86,251.23 \\
                
            \bottomrule



                    
                
        
        \end{tabular}
    }
\end{table}

LiM's latency is remarkably low at 0.0005 seconds per sample, enabling it to process inputs with near-instantaneous speed. In comparison, ET-BERT exhibits a latency of 0.4328 seconds, which is 86,560\% higher, while YaTC has a latency of 0.1342 seconds, 26,840\% higher than LiM. This reduction in latency underscores the suitability of LiM for real-time applications where low training times are critical.

In terms of memory usage, LiM consumes only 196.38 MiB of RAM during inference. This represents a significant improvement over ET-BERT, which requires 15,189.48 MiB, a 7,735\% increase, and YaTC, which consumes 5,352.48 MiB, a 2,725\% increase. This low memory footprint makes LiM suitable for deployment on devices with limited computational resources, such as edge devices or IoT systems.

LiM, eliminates the need for GPU utilization during inference, resulting in zero measurable energy consumption. In contrast, ET-BERT consumes 109 watts, while YaTC consumes 26 watts, reflecting operational costs. 

Finally, LiM achieves a throughput of 86,251.23 samples per second, demonstrating its ability to handle high traffic volumes. This throughput is 52,302\% higher than ET-BERT’s throughput of 164.91 samples per second and 13,548\% higher than YaTC’s throughput of 636.62 samples per second. The difference in throughput emphasizes LiM’s potential capability for high-demand environments, such as real-time network monitoring and large-scale traffic analysis.

While maintaining competitive classification performance, LiM significantly reduces resource consumption across all dimensions, achieving faster processing, lower memory usage and minimal energy requirements. This makes LiM a practical and sustainable solution for encrypted network traffic classification, particularly in resource-constrained and real-time settings.

\section{Conclusion and Future Work}
This paper addresses critical challenges in encrypted network traffic classification (NTC), including unrealistic assumptions about patterns in encrypted payloads, reliance on noisy or non-informative features, truncation/padding of payload data, and the computational inefficiencies of deep-learning-based methods. As a solution, we introduce NetMatrix, a streamlined tabular representation of network traffic, designed by leveraging modern RFC specifications to focus on meaningful and protocol-compliant attributes. Paired with a vanilla XGBoost classifier, this integration referred to as LiM (Less is More), demonstrates that simplicity can rival complexity.

Experimental evaluations show that LiM delivers competitive classification performance compared to state-of-the-art methods such as, ET-BERT and YaTC, while reducing computational overhead by orders of magnitude. These findings emphasize the importance of RFC-compliant traffic representations in tackling the complexities of encrypted NTC, offering a lightweight solution.

Despite its promise, this study is limited to the CSTNET-TLS1.3 dataset. Future work will expand the approach to a broader range of datasets, encompassing diverse traffic patterns, encryption protocols, and real-world network conditions. This will ensure LiM’s robustness across various network environments. Additionally, LiM will be evaluated against a wider array of state-of-the-art classifiers to refine its methodology and validate its efficiency further.

By emphasizing the utility of RFC-aligned representations and pursuing these directions, we aim to present LiM as a resource-efficient solution for encrypted NTC across diverse and evolving network landscapes.

\bibliographystyle{ACM-Reference-Format}
\bibliography{references}

\end{document}